\documentclass[
reprint,
 aps,
 superscriptaddress,
bibnotes,
pra
]{revtex4-2}
\usepackage{physics}
\usepackage{dsfont}
\usepackage{color,newfloat}
\usepackage{graphicx}
\usepackage{dcolumn}
\usepackage{bm}
\usepackage{amsmath,amssymb,amsthm}
\usepackage{mathtools}
\usepackage{multirow}
\usepackage{hhline}
\usepackage{comment}
\usepackage[dvipsnames]{xcolor}
\usepackage{hyperref}
\hypersetup{
    colorlinks=true,
    linkcolor=Maroon,
    citecolor=Maroon,
    urlcolor=Maroon
}
\usepackage{gensymb} % for \degree degree symbol

\usepackage[caption = false]{subfig}
\usepackage{lineno}
\usepackage{siunitx}
\usepackage[mathscr]{euscript}
 \let\mathscr\relax

\usepackage{makecell}

\begin{document}

%%%%%%%%%%%%%%%%%%%%%%%%%%%%%%%%%%%%%%%%%%%%
%%%%%%%%    Title/Authors/Abstract    %%%%%%
%%%%%%%%%%%%%%%%%%%%%%%%%%%%%%%%%%%%%%%%%%%%

\title{
Chiral exceptional bound states in the continuum: a higher-order singularity for on-chip control of quantum emission}

\author{Jin Li}
\affiliation{School of Integrated Circuits, Harbin Institute of Technology (Shenzhen), Shenzhen 518055, China}

\author{Kexun Wu}
\affiliation{School of Integrated Circuits, Harbin Institute of Technology (Shenzhen), Shenzhen 518055, China}

\author{Qi Hao}
\affiliation{Key Laboratory of Quantum Materials and Devices of Ministry of Education, School of Physics, Southeast University, Nanjing 210000, China}

\author{Yan Chen}
\thanks{chenyan@nudt.edu.cn}
\affiliation{Institute for Quantum Science and Technology, National University of Defense Technology, Changsha 410073, China}

\author{Jiawei Wang}
\thanks{wangjw7@hit.edu.cn}
\affiliation{School of Integrated Circuits, Harbin Institute of Technology (Shenzhen), Shenzhen 518055, China}

%%%%%%%%%%%%%%%%%%%%%%%%%%%%%%%%%%
%%%%%%%%     Abstract    %%%%%%%%%
%%%%%%%%%%%%%%%%%%%%%%%%%%%%%%%%%%
\begin{abstract}
We demonstrate a fully integrable and reconfigurable platform for controlling quantum emission by harnessing chiral exceptional bound states in the continuum (BICs) as a higher-order non-Hermitian singularity. Our architecture employs dual-microring resonators evanescently coupled to two waveguides, supporting symmetry-protected BICs. By integrating {a waveguide-coupled reflector} coupled with one resonator as a unidirectional feedback, a pair of orthogonal BICs gets transformed into a single, {chiral quasi-BIC} residing on an exceptional surface. The phase terms in external coupling and inter-modal coupling serve as two independent tuning knobs, enabling unprecedented dynamic control over the spontaneous emission dynamics of individual quantum emitters, including the Purcell enhancement and the emission lineshape. The efficiency in reconfiguring the output intensity gets promoted by more than a factor of two compared to alternative schemes, offering a promising path toward high-speed quantum optical switches and active lifetime control in integrated quantum photonic circuits.
\end{abstract}

\date{\today}

\maketitle

%%%%%%%%%%%%%%%%%%%%%%%%%%%%%%%%%%%
%%%%%%%%     Main text    %%%%%%%%%
%%%%%%%%%%%%%%%%%%%%%%%%%%%%%%%%%%%

%
%
In open physical systems, the intriguing features of non-Hermitian singularities shed profound light on light–matter interactions across classical and quantum domains \cite{RN3, RN2, RN1}. Among these, a hallmark of such systems is the emergence of exceptional points (EPs), singularities where two or more eigenvalues and their corresponding eigenstates coalesce simultaneously, resulting in a collapse of the eigenspace's dimensionality \cite{RN4, RN5, APL}. Furthermore, an equally captivating type of singularity in a non-Hermitian system is the “bound states in the continuum” (BICs) \cite{RN46, RN6}, describing non-radiating resonances embedded within a continuum of radiating waves \cite{RN7}. {In recent explorations using optical microcavities, both of these peculiar singularities have been investigated for tailoring spontaneous emission in cavity quantum electrodynamics (cQED) systems \cite{RN47, RN10, RN31,PRLEP,oeep}, opening new pathways for the on-demand generation and manipulation of single photons as robust information carriers in quantum technologies \cite{RN13, RN11, RN14, RN12,PRBEP}.}

Previous investigations of non-Hermitian cQED in solid-state platforms mainly employed conventional micro- and nanoresonator architectures \cite{RN17, RN15, RN16}, such as micropillars \cite{RN18}, circular Bragg gratings \cite{RN19}, photonic crystals \cite{RN22, RN21, RN20}, and Fabry-Perot cavities \cite{RN23, RN24}, valued for their small mode volumes and high quality factors. Meanwhile, rapid progress in quantum photonic circuits is driving a shift toward integrated architectures. In this context, microring resonators arise as a promising alternative, offering not only inherent compatibility with on-chip, low-loss photonic elements \cite{RN26}, but also exceptional post-fabrication tunability, enabling dynamic reconfiguration of non-Hermitian parameters for quantum emission control \cite{RN29, RN30, RN28}. A particularly elegant strategy for operating a microring system involves the integration of an external reflector, which deterministically couples one propagating mode (e.g., clockwise, CW) into its counter-propagating counterpart (counter-clockwise, CCW), thereby generating a chiral exceptional surface (ES) \cite{RN27, RN34, RN31, RN33}. Recently, such chiral ESs has been exploited to engineer spontaneous emission \cite{RN36, RN37, RN35, NN}. Despite these advances, tunability in key non-Hermitian parameters—particularly radiation through the continuum \cite{RN9, RN30, RN38}, remains largely unexplored in integrated cQED systems.

In this Letter, we demonstrate exquisite control of spontaneous emission from individual quantum emitters by engineering a {chiral exceptional quasi-BIC} in a dual-microring resonator system. By tailoring the phase accumulation within the bus waveguides coupled to two resonators, the radiation channels can be selectively enhanced or suppressed, ideally forming BICs. Concurrently, unidirectional mode coupling via an integrated reflector converts a pair of orthogonal BICs into a chiral one residing on an ES. While coupled to a single quantum emitter (QE), the system exploits two distinct non-Hermitian effects as independent knobs, enabling tunable enhancement of spontaneous emission and dynamic reshaping of its lineshape. Furthermore, for scalable quantum photonic circuits, efficient reconfiguration of both the Purcell factor and output intensity can be achieved with minimal phase tuning near a chiral exceptional quasi-BIC, thereby offering a promising route toward high-speed, on-chip quantum optical switches and active lifetime control of quantum emitters.

\begin{figure*}[]
\includegraphics[width=\linewidth]{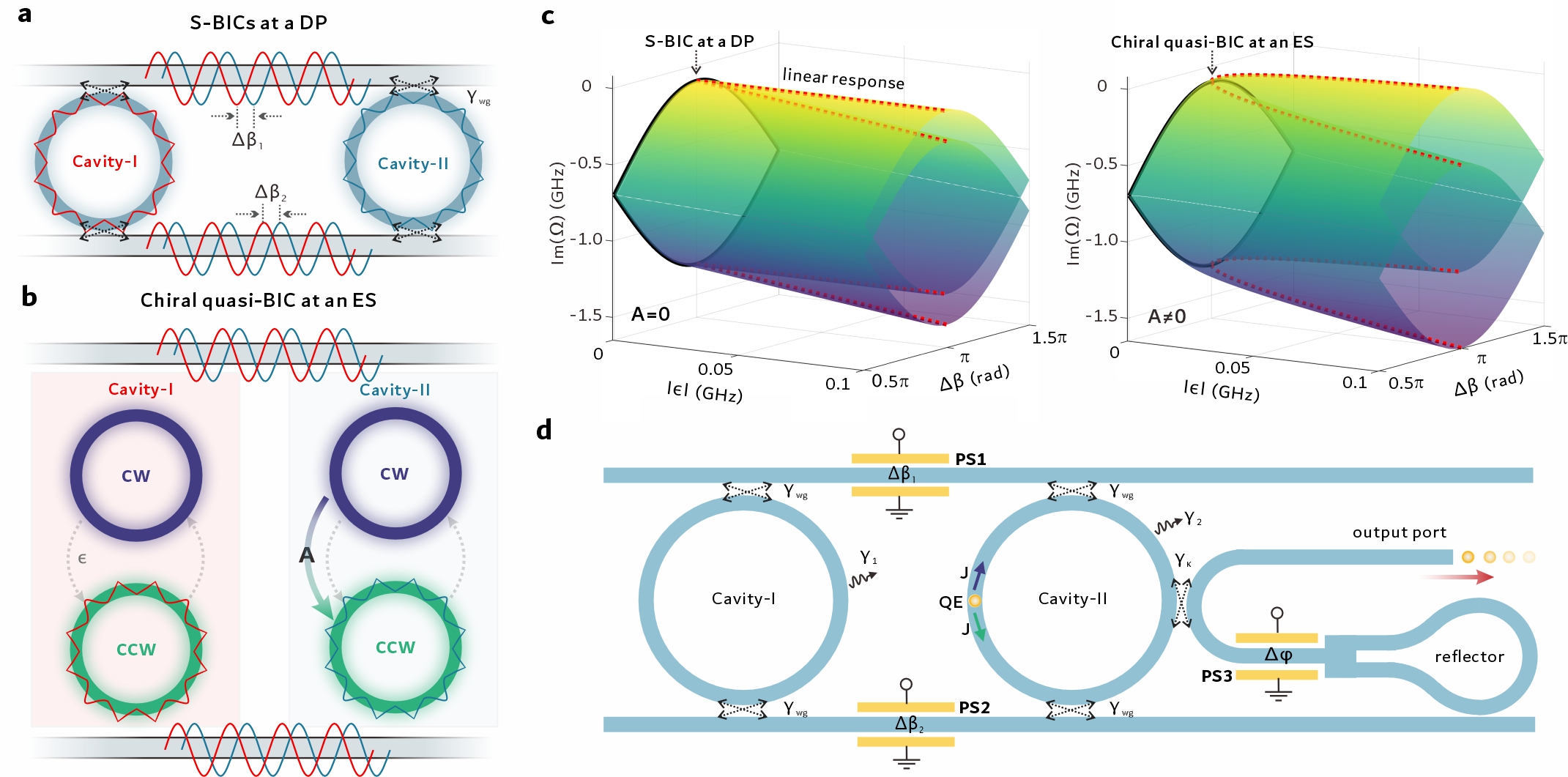}
\caption{(a) Schematic of S-BICs in a dual-ring resonator system. (b) Formation of a chiral exceptional quasi-BIC by introducing a unidirectional coupling term \textbf{\textit{A}}. (c) Numerically calculated imaginary parts of the eigenvalue surfaces in the $\varepsilon$--$\Delta\beta$ parameter space for the S-BIC (left) and chiral exceptional quasi-BIC (right) systems. 
Here, we set $\gamma_{1} = \gamma_{2} + \gamma_{\kappa} = 0.2~\mathrm{GHz}$, $\gamma_{\mathrm{wg}} = 0.4~\mathrm{GHz}$, and $\Delta\omega = 0$. (d) Realization of a circuit-level cQED system harnessing a chiral exceptional quasi-BIC, including a QE embedded in cavity~II and three phase shifters (PS1, PS2, and PS3). 
The QE couples to both CW and CCW modes with an identical electric dipole moment.}
\label{fig1}
\end{figure*}

The discussion starts with a symmetric dual-microring resonator architecture, as illustrated in Fig.~1(a). Each of the two microring resonators supports two spectrally degenerate modes~\cite{RN4}, namely, the CW and CCW propagating modes, and they are not directly coupled. Instead, their interaction is mediated exclusively through two parallel bus waveguides. Since the out-coupled resonances in the two cavities through identical waveguides would interfere with each other, external coupling emerges as a critical degree of freedom. By adopting a traveling-wave basis $\left[ a_{\mathrm{CCW}},\, a_{\mathrm{CW}},\, b_{\mathrm{CCW}},\, b_{\mathrm{CW}} \right]^{\mathrm{T}}$, where \textbf{\textit{a}} and \textbf{\textit{b}} correspond to the fields in the two cavities (see SM Note~S1), the system can be described using the Hamiltonian as follows:
\begin{equation}
H=\left( \begin{matrix}
   {{\Omega }_{1}} & 0 & {{\Gamma }_{1}} & 0  \\
   0 & {{\Omega }_{1}} & 0 & {{\Gamma }_{2}}  \\
   {{\Gamma }_{2}} & 0 & {{\Omega }_{2}} & 0  \\
   0 & {{\Gamma }_{1}} & 0 & {{\Omega }_{2}}  \\
\end{matrix} \right),\
\end{equation}
where ${{\Omega }_{1}}={{\omega }_{1}}-i{{\gamma }_{1}}-i2{{\gamma }_{\text{wg}}}$, ${{\Omega }_{2}}={{\omega }_{2}}-i{{\gamma }_{2}}-i2{{\gamma }_{\text{wg}}}$, \({{\Gamma }_{1}}=-i2{{\gamma }_{\text{wg}}}{{e}^{i\Delta {{\beta }_{1}}}}, {{\Gamma }_{2}}=-i2{{\gamma }_{\text{wg}}}{{e}^{i\Delta {{\beta }_{2}}}}\), $\omega_{1}$ ($\omega_{2}$) and $\gamma_{1}$ ($\gamma_{2}$) denote the resonance frequency and intrinsic cavity loss rate of the two cavities, respectively. $\gamma_{\mathrm{wg}}$ is the coupling rate between the resonators and the two bus waveguides, and $\Delta\beta_{1}$ ($\Delta\beta_{2}$) represent the relative phase shifts between the two interfering light waves for the two bus waveguides serving as radiation channels, respectively.

{The phase terms between the two radiative components, $\Delta\beta_1$ and $\Delta\beta_2$, serve as effective parameters for regulating the external coupling. When the two resonance frequencies are degenerate and $\Delta\beta_1 = \Delta\beta_2 = \pi/2$, the radiation fields interfere constructively through the waveguides, resulting in maximized radiation leakage into the continuum. In contrast, when the phase shifts satisfy the destructive interference condition ($\Delta\beta_1 = \Delta\beta_2 = \pi$), the outgoing radiation fields cancel each other, effectively eliminating the waveguide-induced leakage channels. Under this condition, the system supports a symmetry-protected bound state in the continuum (S-BIC) with ideally vanishing radiative loss and an infinite radiative quality (Q) factor.} Notably, the normalized eigenvectors $\left[1,\,0,\,1,\,0\right]^{\mathrm{T}}$ and $\left[0,\,1,\,0,\,1\right]^{\mathrm{T}}$ reveal opposite mode chiralities and share a spectral degeneracy at a diabolic point (DP).

Here, by introducing a unidirectional coupling between the CW and CCW components (Fig. 1(b)), the original orthogonality between the pair of S-BICs is broken. The system Hamiltonian is consequently modified to:
\begin{equation}
H=\left( \begin{matrix}
   {{\Omega }_{1}} & 0 & {{\Gamma }_{1}} & 0  \\
   0 & {{\Omega }_{1}} & 0 & {{\Gamma }_{2}}  \\
   {{\Gamma }_{2}} & 0 & {{\Omega }_{2}}^{'} & A  \\
   0 & {{\Gamma }_{1}} & 0 & {{\Omega }_{2}}^{'}  \\
\end{matrix} \right),\
\end{equation}
where ${{\Omega }_{2}}^{'}={{\omega }_{2}}-i{{\gamma }_{2}}-i{{\gamma }_{\kappa }}-i2{{\gamma }_{\text{wg}}}$, \textbf{\textit{A}} specifies the coupling coefficient applied to the cavity-II (from the CW to the CCW mode in this case), and $\gamma_{\kappa}$ accounts for the additional loss induced by the unidirectional coupling. Assuming a symmetric design of the two bus waveguides for simplicity (i.e., $\Delta\beta_{1} = \Delta\beta_{2} = \Delta\beta$), the eigenfrequencies and eigenvectors can be obtained as: \({{\Omega }_{\pm }}=\frac{{{\Omega }_{1}}+{{\Omega }_{2}}^{'}}{2}\pm \sqrt{{{\left( \frac{{{\Omega }_{1}}-{{\Omega }_{2}}^{'}}{2} \right)}^{2}}+4{{e}^{i2\Delta \beta }}{{\gamma }_{\text{wg}}}^{2}}\), ${{v}_{\pm }}={{\left( \frac{{{\Omega }_{1}}-{{\Omega }_{2}}^{'}}{2}\pm \sqrt{{{\left( \frac{{{\Omega }_{1}}-{{\Omega }_{2}}^{'}}{2} \right)}^{2}}+4{{e}^{i2\Delta \beta }}{{\gamma }_{\text{wg}}}^{2}},0,2{{e}^{i\Delta \beta }}{{\gamma }_{\text{wg}}},0 \right)}^{T}}$. 

The system’s chirality becomes deterministic, with a purely CCW component existing. Here, different higher-order singularities can be engineered. A fourth-order EP can be reached by introducing $\Delta\omega = \omega_{1} - \omega_{2} = 2\gamma_{\mathrm{wg}}$ (see SM Note~2). For identical resonance frequencies ($\Delta\omega = 0$), a chiral exceptional quasi-BIC, as a peculiar type of higher-order singularity, can be established, featuring minimized radiation loss {(practically limited by the presence of $\gamma_{1}$, $\gamma_{2}$, and $\gamma_{\kappa}$)}.

In practice, fabrication imperfections (e.g., surface roughness) may randomly perturb the system and induce backscattering between the CW and CCW components. For simplicity, the perturbation Hamiltonian can be expressed as
\(
H_{\mathrm{perturb}} =
\begin{pmatrix}
\varepsilon & \varepsilon & 0 & 0 \\
\varepsilon & \varepsilon & 0 & 0 \\
0 & 0 & \varepsilon & \varepsilon \\
0 & 0 & \varepsilon & \varepsilon
\end{pmatrix},
\) where $\varepsilon$ quantifies the perturbation strength applied uniformly to both microresonators.

Figure~1(c) summarizes the numerically calculated imaginary parts of the eigenvalue surfaces in the $|\varepsilon|$--$\Delta\beta$ parameter space for the two systems. For the conventional dual-ring system, when $|\varepsilon| = 0$, the radiation is regulated by $\Delta\beta$. The two curves correspond to the spectrally degenerate long- and short-lived modes, representing two DPs. As perturbations are introduced, four Riemann sheets reveal mode splitting with an amplitude linearly proportional to $|\varepsilon|$. In contrast, for the proposed system in Fig.~1(b), the two curves at $|\varepsilon| = 0$ essentially lie on the corresponding second-order ESs, where eigenvalues and eigenvectors coalesce simultaneously. Under perturbation, the eigenvalues exhibit a nonlinear enhancement proportional to $\sqrt{\varepsilon}$, directly reflecting the signature of an EP.

We now employ the peculiar chiral exceptional quasi-BIC shown in Fig.~1(b) for the regulation of on-chip integrated quantum sources. As illustrated in Fig.~1(d), a QE is located within the mode volume of cavity~II. A waveguide integrated with a reflector (e.g., a Sagnac interferometer) provides the requisite non-Hermitian coupling, \(\mathbf{A} = -2i\gamma_{\kappa} \left| r \right| \exp(-i\Delta\varphi)\), where $r$ is the field reflection amplitude and $\Delta\varphi$ is the phase variation. The full Hamiltonian of this cQED system is detailed in SM Note~S3. For semiconductor quantum dots with typical free-space lifetimes of $\sim$1--10~ns and integrated microrings offering photon lifetimes of $\sim$0.1--1~ns, the system operates in the weak-coupling regime~\cite{RN35,RN31}, where an emitted photon escapes before re-coupling to the emitter. Under this approximation, the dipole–cavity dynamics is governed by:
\begin{equation}
    i\frac{d}{dt}
\begin{pmatrix}
   a_{\text{CCW}} \\
   a_{\text{CW}} \\
   b_{\text{CCW}} \\
   b_{\text{CW}} \\
\end{pmatrix}
= H
\begin{pmatrix}
   a_{\text{CCW}} \\
   a_{\text{CW}} \\
   b_{\text{CCW}} \\
   b_{\text{CW}} \\
\end{pmatrix}
+ J
\begin{pmatrix}
   0 \\
   0 \\
   e^{-i(\omega_{\text{QE}} t + \phi)} \\
   e^{-i(\omega_{\text{QE}} t - \phi)} \\
\end{pmatrix},
\end{equation}
where $\omega_{\mathrm{QE}}$ is the emission frequency of the quantum emitter, $J$ is the coupling constant for optical fields at a particular transverse position along the propagation direction for a traveling-wave resonance, and the phase term $\phi$ accounts for the accumulated dynamical phase between the emitter and a reference point (e.g., the waveguide–ring junction). The steady-state response can be obtained as: \({{\left( {{a}_{CCW}},{{a}_{CW}},{{b}_{CCW}},{{b}_{CW}} \right)}^{T}}=G({{\omega }_{QE}}){{\left( 0,0,J{{e}^{-i\phi }},J{{e}^{i\phi }} \right)}^{T}}\), where $G(\omega_{\mathrm{QE}}) = (\omega_{\mathrm{QE}} I - H)^{-1}$ is the system's Green function (see SM Note~S3)~\cite{RN41,RN40,RN42}.

\begin{figure}[]
\includegraphics[width=\columnwidth]{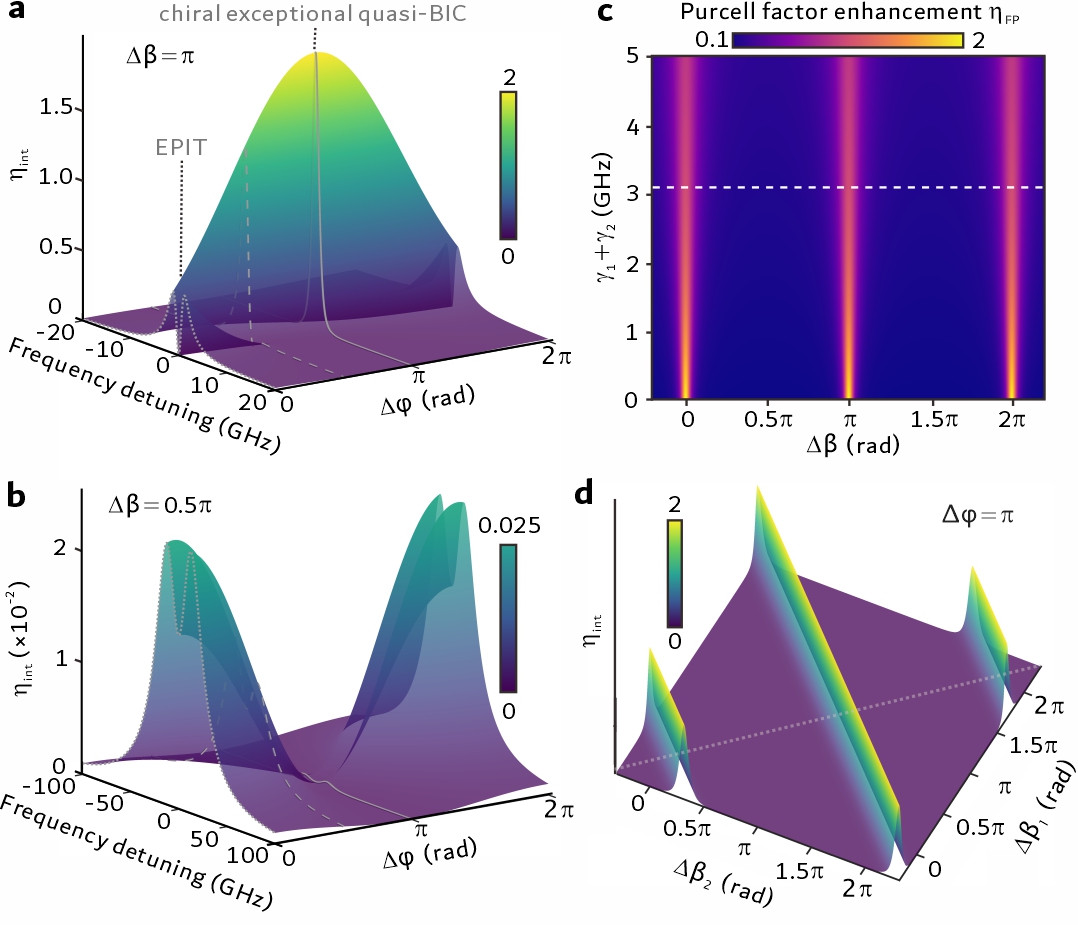}
\caption{(a--b) Calculated $\eta_{\mathrm{int}}$ versus frequency and phase difference $\Delta\varphi$ for $\Delta\beta = \pi$ (a) and $\Delta\beta = \pi/2$ (b). Here, we set $\Delta\omega = 0$, $\left| r \right|=1$, $\gamma_{\kappa} = 2.5~\mathrm{GHz}$, $\gamma_{\mathrm{wg}} = 5~\mathrm{GHz}$, and $\gamma_{1} = \gamma_{2} = 0~\mathrm{GHz}$. (c) Calculated $\eta_{\mathrm{FP}}$ as a function of $\Delta\beta$ and the sum of the intrinsic cavity losses, $\gamma_{1} + \gamma_{2}$. The dashed line denotes the case for a typical TFLN-based microring resonator with an intrinsic $Q$ factor of 20000. (d) Calculated $\eta_{\mathrm{int}}$ as a function of $\Delta\beta_{1}$ and $\Delta\beta_{2}$ at zero frequency detuning and $\Delta\varphi = \pi$. The dashed line denotes the ideal symmetric external coupling case where $\Delta\beta_{1} = \Delta\beta_{2}$.}
\label{fig2}
\end{figure}

Here, we analyze the intensity collected at the remaining output port of the coupled waveguide (see Fig.~1(d)), which can be expressed as
\begin{equation}
I_{\mathrm{output}} = \left| \sqrt{2\gamma_{\kappa}} \left( b_{\mathrm{CW}} \left| r \right| \ e^{i\Delta\varphi} + b_{\mathrm{CCW}} \right) \right|^{2},
\end{equation}
where
\begin{align}
b_{\mathrm{CCW}} &= \frac{-(\Omega_{1} - \omega_{\mathrm{QE}}) J e^{i\phi}}{(\Omega_{1} - \omega_{\mathrm{QE}})(\Omega_{2}^{'} - \omega_{\mathrm{QE}}) - \Gamma_{1}\Gamma_{2}} \notag \\ 
&\quad + \frac{J e^{-i\phi} A (\Omega_{1} - \omega_{\mathrm{QE}})^{2}}{\left[ (\Omega_{1} - \omega_{\mathrm{QE}})(\Omega_{2}^{'} - \omega_{\mathrm{QE}}) - \Gamma_{1}\Gamma_{2} \right]^{2}},
\end{align}
\begin{equation}
b_{\mathrm{CW}} = \frac{-(\Omega_{1} - \omega_{\mathrm{QE}}) J e^{-i\phi}}{(\Omega_{1} - \omega_{\mathrm{QE}})(\Omega_{2}^{'} - \omega_{\mathrm{QE}}) - \Gamma_{1}\Gamma_{2}}.
\end{equation}

For simplicity, the discussion adopts the approximation of negligible intrinsic backscattering ($\varepsilon = 0$, see discussions on the influence of backscattering in SM Note~S4). Two phase terms, $\Delta\beta_{1}$ and $\Delta\beta_{2}$, are simultaneously tuned to an identical value $\Delta\beta$. To quantify the output spectral intensity at the port, we define a relative spontaneous emission enhancement factor
\(\eta_{\mathrm{int}} = \frac{I_{\mathrm{output}}}{\max(I_{\mathrm{DP}})},\) where $I_{\mathrm{DP}}$ is the total output intensity collected from a conventional waveguide-coupled microring within a cQED system (see SM Note~S5). Figure~2(a) presents the evolution of $\eta_{\mathrm{int}}$ when the system is configured to a chiral exceptional quasi-BIC ($\Delta\beta = \pi$). The emission lineshape is efficiently modified upon varying $\Delta\varphi$. The maximal intensity is obtained at $\Delta\varphi = \pi$, exhibiting a squared-Lorentzian profile in sharp contrast to the regular Lorentzian profile in the conventional DP case. At $\Delta\varphi = 0$, the resulting spectrum is doublet-shaped with a zero value at the central frequency. This ``exceptional-point-induced transparency (EPIT)'' phenomenon arises from perfect destructive interference between the CCW and back-reflected components. When the system is tuned away from the chiral exceptional quasi-BIC ($\Delta\beta = \pi/2$, Fig.~2(b)), the reinforced external coupling via the two bus waveguides does not lead to a homogeneous suppression of the spectra upon varying $\Delta\varphi$. Instead, the minimal emission intensity is reached at $\Delta\varphi = \pi$, which is only $\sim 0.01\%$ of the maximized value in Fig.~2(a).

To quantitatively elucidate the cavity enhancement within the weak coupling regime, the local density of optical states and the Purcell enhancement factor can be expressed as \cite{RN47, RN10, RN42}:
\begin{equation}
\
\rho_{\mathrm{LDOS}}(\omega_{\mathrm{QE}}) = -\frac{2}{\hbar \varepsilon_{0}} \mathbf{d} \, \operatorname{Im} \big[ G(\omega_{\mathrm{QE}}) \big] \mathbf{d}^{*},
\
\end{equation}
\begin{equation}
F_{p}(\omega_{\mathrm{QE}}) = 1 + \frac{\rho_{\mathrm{LDOS}}(\omega_{\mathrm{QE}})}{\rho_{\mathrm{hom}}},
\end{equation}
where $\mathbf{d}$ denotes the dipole moment vector of the QE (in this case, $\mathbf{d} = [0,\,0,\,1,\,1]^{\mathrm{T}}$), and $\rho_{\mathrm{hom}}$ represents the LDOS in a homogeneous medium. {Using Eqs. (7) and (8), under the same parameter configuration as in Figs. 2(a)–(b), the calculated Purcell factor at the resonant frequency oscillates between approximately 500 ($\Delta\beta = \pi/2$) and 8000 ($\Delta\beta = \pi$) as $\Delta\beta$ varies. It should be noted that this prediction corresponds to an idealized scenario. In practice, the achievable Purcell factor is expected to be substantially lower due to several non-ideal effects, including the dipole emitter’s properties, imperfect alignment between the emitter dipole moment and the cavity mode, nonradiative decay channels, and the finite linewidth of the emitter itself. To better assess the dynamical characteristics, we introduce an enhancement factor of the Purcell factor $\eta_{\mathrm{FP}}$ defined as \(\eta_{\mathrm{FP}} = \frac{F_{P}(\omega_{\mathrm{QE}})}{F_{P,\mathrm{DP}}(\omega_{\mathrm{QE}})},\) where $F_{P,\mathrm{DP}}$ denotes the Purcell factor under the DP condition.}

\begin{figure}[]
\includegraphics[width=\columnwidth]{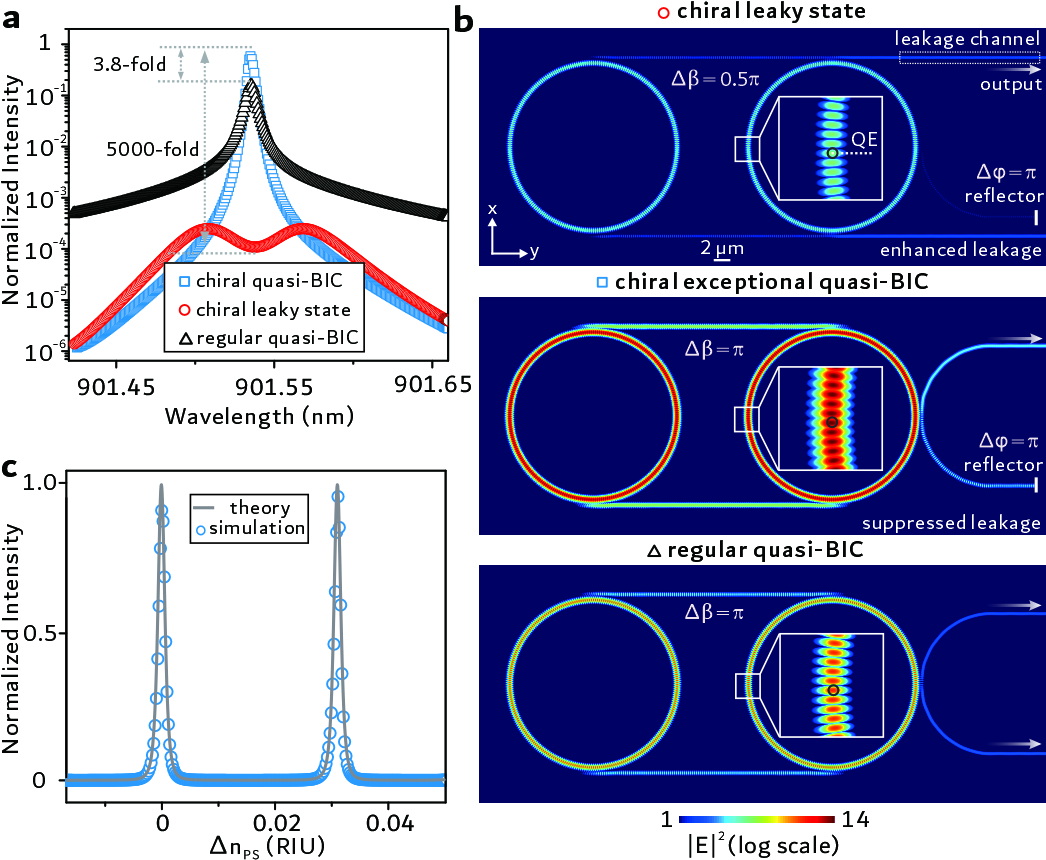}
\caption{(a) Simulated output spectra for the chiral exceptional quasi-BIC ($\Delta\varphi = \pi$, $\Delta\beta = \pi$), chiral leaky state ($\Delta\varphi = \pi$, $\Delta\beta = \pi/2$), and regular quasi-BIC ($\Delta\beta = \pi$). The two identical microrings have a radius of 10~$\mu$m, bus waveguides and rings are 400~nm wide with a 250~nm coupling gap. The feedback waveguide has a 300~nm gap and a perfect electric conductor at its end as a reflector. The QE is modeled as a point dipole mimicking a hybrid InGaAs quantum dot at $\sim$901~nm. (b) Mode electric field intensity $|E|^2$ on a logarithmic scale for the three cases in (a). Insets: zoomed-in views of the region surrounding the embedded QE. (c) Extracted output normalized intensity versus $\Delta n_{PS}$ applied to the two phase shifters integrated on the bus waveguides. Solid line: theoretical fit.}
\label{fig3}
\end{figure}

Figure~2(c) presents the calculated $\eta_{\mathrm{FP}}$ as a function of $\Delta\beta$ and the total intrinsic cavity losses, $\gamma_{1} + \gamma_{2}$. $\eta_{\mathrm{FP}}$ reaches a maximum of 2 at an optimally configured chiral exceptional quasi-BIC and decreases as the system moves away from this singularity. The upper bound of $\eta_{\mathrm{FP}}$ is fundamentally constrained by the intrinsic cavity losses. The experimental configuration, featuring two integrated phase shifters for the independent control of $\Delta\beta_{1}$ and $\Delta\beta_{2}$, enables dynamic reconfiguration of the emission dynamics. As revealed in Fig.~2(d), by keeping one phase constant, both the output intensity and $F_P$ can be efficiently tuned while preserving near-optimal enhancement close to the chiral exceptional quasi-BIC (see SM Note~S6 for details). This flexibility underscores the potential for on-chip reconfiguration of quantum emission dynamics.  

Beyond the analytical model, we further investigate a realistic implementation of a {chiral exceptional quasi-BIC} on a thin-film lithium niobate (TFLN) integrated photonic platform, notable for hybrid integration with semiconductor quantum dots as an on-chip quantum source~\cite{RN34,RN43}. {Our numerical simulation was conducted using a two-dimensional finite-element method (COMSOL Multiphysics). Scattering boundary conditions were applied at the outermost boundaries to avoid spurious reflections. A locally refined, non-uniform triangular mesh was employed in the resonator region, with a mesh size below 50 nm to ensure numerical accuracy. For the refractive indices, we used a constant value of  2 for LN waveguides and 1.55 for the silica cladding layer. Material dispersion was neglected, given the narrow spectral range considered. The quantum emitter was modeled as an in-plane oriented point magnetic dipole. The dipole moment was aligned along the in-plane direction (1,0) in the (x,y) coordinate system and positioned at the center of the microring waveguide cross-section, as schematically illustrated in Fig. 1(d). The output intensity and emission lineshape were extracted by performing an integration of the time-averaged Poynting vector across the width of the waveguide output port.}

As shown in Fig.~3(a), the squared-Lorentzian lineshape at $\Delta\beta = \pi$ and the mode-splitting doublet feature at $\Delta\beta = \pi/2$ are obtained, producing a colossal intensity contrast of 5000-fold through phase tuning. The distinct external coupling behaviors are further ascertained by the simulated mode field profiles in Fig.~3(b) (top and middle panels). In comparison, the regular dual-ring system (Fig.~3(b), bottom) demonstrates that the regular quasi-BIC exhibits a 3.8-fold reduction in output intensity relative to the chiral quasi-BIC (see SM Note~S7 for details). By locally tuning the refractive index of the two bus waveguides via integrated phase shifters (Fig.~3(c)), the output spectral intensity can be fully adjusted between $10^{-4}$ and 1, with the modulation curve closely matching the theoretical model in Fig.~2(d).

\begin{figure}[]
\includegraphics[width=\columnwidth]{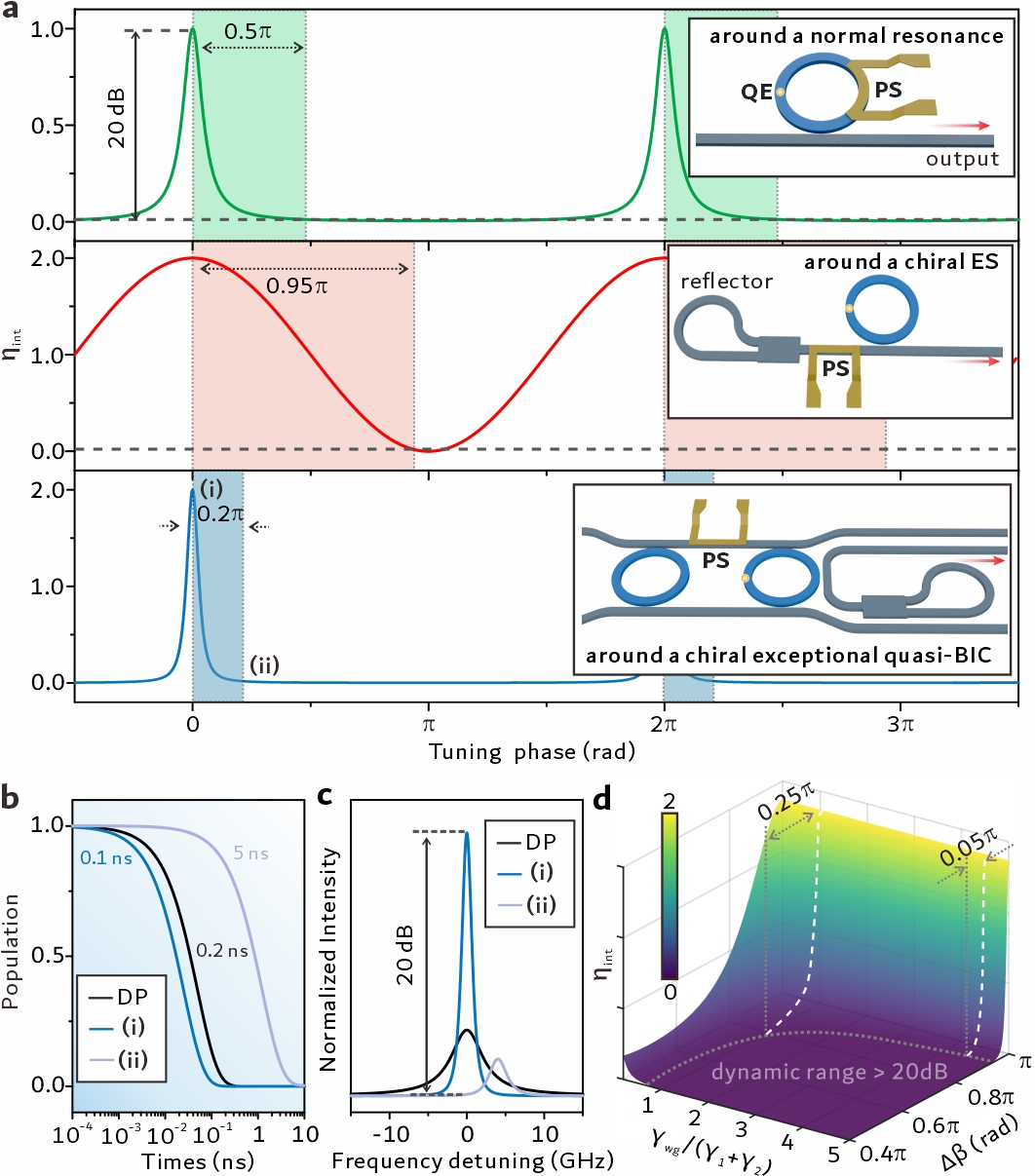}
\caption{(a) Calculated $\eta_{\mathrm{int}}$ versus applied phase tuning onto the corresponding phase shifter in three different configurations, including the conventional microring system operated at a DP (top), the single microring system operated at a chiral ES (middle), and a dual-ring system operated around a chiral exceptional quasi-BIC (bottom). Insets: corresponding system schematics. (b) Calculated lifetime traces of a single QE upon reconfiguration of $\Delta\beta$. (c) Calculated emission spectra upon reconfiguration of $\Delta\beta$. The cases under a DP condition are also presented in (b--c) for comparison. (d) Map of $\eta_{\mathrm{int}}$ as a function of $\Delta\beta$ and the loss contrast $\gamma_{\mathrm{wg}}/(\gamma_{1} + \gamma_{2})$. The dashed lines indicate the phase amplitude required for achieving a 20~dB dynamic range of $\eta_{\mathrm{int}}$.}
\label{fig4}
\end{figure}

Both theory and simulations show that operating near a chiral exceptional quasi-BIC enables efficient reconfiguration of spontaneous emission. Here, we compare $\eta_{\mathrm{int}}$ reconfiguration with two previously explored schemes. In a conventional microring-resonator-based cQED system, reconfiguration can be performed via resonance detuning between the resonant mode and $\omega_{\mathrm{QE}}$. Considering practical intrinsic cavity losses of $\sim 15~\mathrm{GHz}$ (corresponding to an intrinsic $Q$ factor of 20000 \cite{RN43}), a 20~dB dynamic range in $\eta_{\mathrm{int}}$ requires a phase tuning amplitude of $\sim 0.5\pi$ (Fig.~4(a) top). Notably, by engineering the resonant mode from a common DP to a chiral EP, the maximal $\eta_{\mathrm{int}}$ is enhanced by two-fold (Fig.~4(a) middle). An equivalent dynamic reconfiguration can now be achieved by steering the system towards the vicinity of EPIT, requiring phase tuning of $\Delta\varphi \sim 0.95\pi$ on the feedback arm of the reflector.  

Strikingly, in our proposed system, the maximal $\eta_{\mathrm{int}}$ matches that of the chiral EP scheme. In contrast, the reconfiguration is accomplished by varying the external coupling strength. A 20~dB dynamic range requires tuning one of the two external coupling channels (i.e., $\Delta\beta_{1}$ or $\Delta\beta_{2}$) to $\sim 0.2\pi$ (Fig.~4(a) bottom). Assuming a radiative lifetime of 200~ps in a DP condition as a baseline, such tuning facilitates reconfiguration of the lifetime between 100~ps and 5~ns (Fig.~4(b)). Moreover, the regulated radiation rate corresponds to distinct emission lineshapes in the spectral domain (Fig.~4(c)).  

This stands in stark contrast to conventional paradigms, where the radiation loss remains static. Our approach transcends this limitation by leveraging multiple degrees of freedom to reconfigure the non-Hermiticity around a chiral exceptional quasi-BIC. Both phase shifters on the two bus waveguides can be utilized with an equal amount of tuning, resulting in a doubling of tuning efficiency. Furthermore, the external coupling loss $\gamma_{\mathrm{wg}}$ can be readily adjusted by changing the gap spacing between the bus waveguides and microrings. Figure~4(d) presents the mapping of $\eta_{\mathrm{int}}$ upon the regulation of the loss ratio between the external coupling and intrinsic cavity, $\gamma_{\mathrm{wg}} / (\gamma_{1} + \gamma_{2})$, and $\Delta\beta$. The mapping reveals that the phase tuning required to achieve a 20~dB dynamic range can be drastically reduced to one-quarter of its original value by simply increasing $\gamma_{\mathrm{wg}}$ by a factor of 4.5.

In summary, we have theoretically demonstrated an integrated cQED system combining two non-Hermitian singularities to achieve precise control over quantum light emission. The introduced {chiral exceptional quasi-BIC} features maximal mode chirality at an EP and meanwhile ideally zero radiation loss. By reconfiguring the system around such a higher-order singularity, the spontaneous emission dynamics, including the emission intensity, lifetime, and spectra lineshape, can be freely manipulated with minimized requirement of phase tuning amplitude in the external coupling. 

Looking forward, this concept is well-suited for various material platforms with integrated QEs, especially TFLN \cite{RN34, RN29, RN28}, silicon nitride \cite{LPRy}, and silicon carbide. {Crucially, the active tuning required for switching can leverage the strong electro-optic effect in platforms such as TFLN, enabling phase modulation with bandwidths exceeding 40 GHz  and switching on nanosecond to picosecond timescales \cite{nptin, optin}. This high-speed capability is directly compatible with deterministic quantum emitters, such as InGaAs quantum dots \cite{LPR}. Their heterogeneous integration with lithium niobate platforms \cite{NC, NM} paves the way for  fully on-chip, dynamically reconfigurable quantum nodes. The demonstrated on-the-fly reconfigurability positions this device as a versatile building block that can be dynamically configured as an on-chip single-photon source, a quantum optical switch, or a long-lived quantum memory \cite{RN45}. Such multifunctionality represents a critical capability for scalable quantum computing \cite{RN26} and communicating architectures.}

\section*{Acknowledgement}
J.W. and Y.C. acknowledges the support from the National Natural Science Foundation of China under Grants 62422503, 12474375 and 12374476, the Guangdong Basic and Applied Basic Research Foundation Regional Joint Fund under Grant 2023A1515011944, Science and Technology Innovation Commission of Shenzhen under Grants JCYJ20220531095604009 and RCYX20221008092907027.

\bibliography{ref.bib}
\end{document}